# Self-aligned multilayered nitrogen vacancy diamond nanoparticles for high spatial resolution magnetometry of microelectronic currents

Yash Gokhale[1,2,*], Brandon S Coventry[2,3,*], Tsani Rogers[1], Maya Lines[4], Anna Vena[1], Jack Phillips[1], Tianxiang Zhu[1], Ilhan Bok[1,2], Dariana J. Troche[5], Mitchell Glodowski[1], Adam Vareberg[1], Suyash Bhatt[1,4], Alireza Ashtiani[1], Kevin W. Eliceiri[1,6,7], Aviad Hai[1,2,8*]

[1]Department of Biomedical Engineering, University of Wisconsin–Madison

[2]Wisconsin Institute for Translational Neural Engineering, University of Wisconsin-Madison

[3]Department of Neurological Surgery, University of Wisconsin-Madison

[4]Department of Neuroscience, Lawrence University

[5]Department of Mechanical Engineering, University of Puerto Rico at Mayagüez

[6]Department of Medical Physics, University of Wisconsin-Madison

[7]Morgridge Institute for Research, Madison, Wisconsin

[8]Department of Electrical and Computer Engineering, University of Wisconsin–Madison

* These authors contributed equally to the work

Address correspondence to AH

1550 Engineering Drive, Rm 2112

Madison, WI, 53706

phone: 608-890-3411

e-mail: ahai@wisc.edu

**Abstract**

Nitrogen Vacancy diamond nanoparticles (NVNPs) are increasingly integrated with methods for optical detection of magnetic resonance (ODMR), providing new opportunities that span the visualization of magnetic fields in microelectronic circuits, environmental sensing and biology. However, only a small number of studies utilize aggregates of NVNPs for surface-wide magnetometry being that spin orientations in aggregate NVNPs are inherently misaligned, precluding their use for proper magnetic field detection compared with expensive monocrystalline diamonds. A postprocessing method for layering NVNPs with aligned NV center orientations can potentially facilitate superior NV magnetometry by allowing sensitive detection combined with simplified probe preparation. We present a novel technology for creating densely stacked monolayers of NVNP with inherent interlayer alignment for sensitive measurement of local magnetic field perturbations in microelectronic traces. We establish spatial characteristics of deposited aggregates and demonstrate their ability to capture magnetic dipoles from conducting microwires via ODMR. Our approach forms a novel accessible protocol that can be used for broad applications in micromagnetometry.

# Main

Nitrogen vacancy (NV) diamond magnetometry harnesses optomagnetic sensitivities of an electron pair near vacancy centers of irradiated diamond substrates to measure femtotesla (fT) level magnetic fields at nanometer scale resolution[1–5]. By relying on optically detected magnetic resonance (ODMR) of electron spectra splitting in proportion to nearby magnetic fields[6–8,5,9] a multitude of applications have been realized in quantum computing[10–14], characterization of magnetic nanostructures[15–18], and investigation of biological and neural magnetism[19–23]. Most studies leverage quantum grade single-crystal bulk diamond with NV-implanted thin layer requiring precision manufacturing[24–26]. Implementing the technology for integrated electronics is even more challenging due to difficulties in adhering diamond to thin film devices or fabricating directly on diamond[27–29]. Alternative architectures of NV nanodiamond probes have been utilized for serial scanning or single point recording by conjugating single NV nanocrystals with cantilever structures[30–33]. These methods provide spatiotemporal sensitivity comparable to mainstay microelectronic magnetic characterization techniques such as superconducting quantum interference devices (SQUIDs) microscopy[34,35] and magnetic force microscopy (MFM)[36,37] but lack the parallel optical readouts available to NV magnetometry. A subset of these new designs rely on hybridization with single NV diamond nanoparticles (NVNPs) that are primarily emerging as highly photostable fluorescent imaging probes[38–40] which have also been utilized for direct magnetomechanical manipulation and detection of biomolecules[17,41]. NVNPs are synthesized by more simplified processes including milling of high pressure high temperature fabricated diamonds, laser ablation, or detonation[26,42] exhibiting reduced purity, NV distribution uniformity issues, and limited coherence times compared with industrially grown slabs[26]. Nonetheless, their intrinsic ability to be integrated with ODMR presents new opportunities in magnetic characterization including visualization of magnetic fields generated by currents in conductive patterned devices. A small number of studies utilized aggregates of NVNPs for surface-wide magnetometry[43,44] but spin orientations in aggregate NVNPs are inherently misaligned, precluding the use of such preparations for coherent determination of magnetic field vectors normally achieved by monocrystalline diamonds. A postprocessing method for layering NVNPs with aligned NV center orientations can potentially facilitate superior NV magnetometry by allowing sensitive detection combined with simplified probe preparation. In this study, we present a novel and reproducible protocol for creating densely stacked monolayers of NVNP with inherent interlayer alignment for sensitive measurement of local magnetic field perturbations induced by microelectronic currents. We establish spatial characteristics of deposited aggregates and demonstrate their ability to capture magnetic dipoles from conducting microwires via ODMR. We then generate NVNP-mediated visualization maps of currents generated by microfabricated 100 nm Ti/Au traces with single micrometer spatial resolution. Using electron microscopy analysis and computational methods we explain the mechanism of detection by predicting a level of up to 20 % alignment across multiple tiers of the multilayer NVNP preparation required to achieve sensitivities measured experimentally. Our approach forms a novel accessible protocol that can be used for broad applications in micromagnetometry.

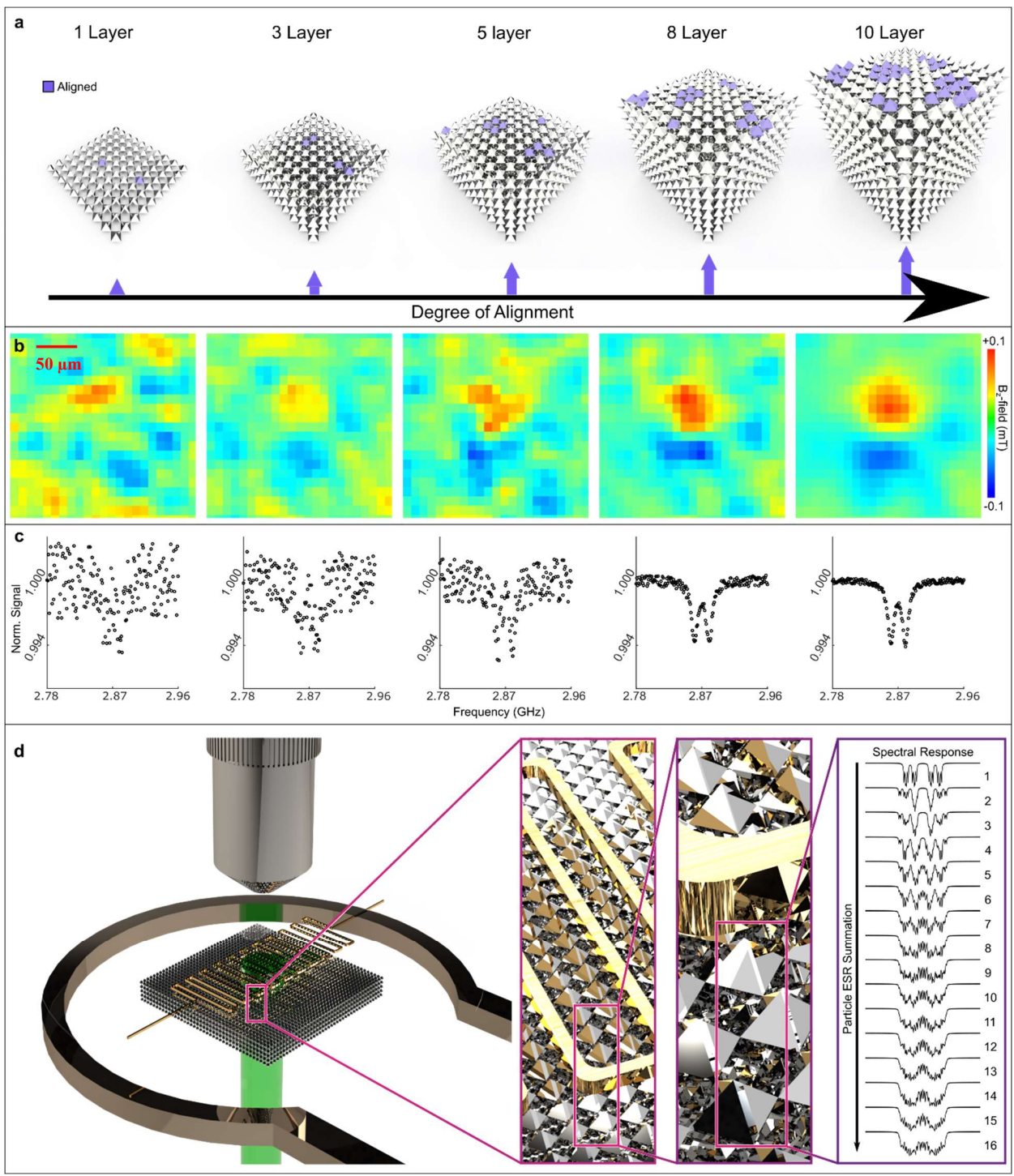

**Fig. 1: Increased alignment of multilayered nitrogen vacancy diamond nanoparticles (NVNP) relying on nucleation sites from previous layers to promote improved sensitivity to magnetic fields.** (a) Depiction of aggregates of aligned particles with increasing coverage coinciding with increasing layer count. (b) Sensitive ODMR map of magnetic flux profile of a dipole emerging from the surface of a 100 × 100 µm², 30 nm, thin film iron oxide probed by multiple layers of NVNPs (rightmost panel – 30 layers) and decreasing in sensitivity to a randomly aligned single NVNP layer (leftmost panel). Scale bar 50 µm. (c) Increased Layering corresponds with decreased MW spectral noise. (d) Schematic of magnetometry setup (not to scale) for measuring on-chip charge fluctuations using multilayered NVNPs. A 532 nm excitation pulse is applied to particles from below with fluorescence collected at 637 nm by objective above sample, with MW (2.78 – 2.96 GHz) applied by a printed circuit antenna proximal to samples during current propagation. Right: particle ensemble electron spin resonance (ESR) profiles calculated over increasing layer count.

*Incremental layering of NVNPs increases performance*

Interparticle interactions within solution milieu during solvent evaporation govern bottom-up nanoparticle self-assembly into highly tunable ordered superlattices, generating unique structural alignment and functionality[45]. Self-aligned superlattice diamond-NV sensors fabricated by polydimethylglutarimide (PMGI) wet etching of nanopatterned pillars[46], microwave (MW) plasma chemical vapor deposition (CVD) of umbrella-shaped patterned protrusions[47], or functionalization through amine-reactive cross linking[48] show vast improvements in sensor sensitivity but require highly involved nanolithography. Other powerful examples include spin-wave electronic sensors probed using ODMR by exploiting randomly scattered nanodiamonds on patterned substrates[49], and single nanodiamonds levitated in vacuum to greatly improve sensitivity and readout resolution[50]. But these efforts also require specialized techniques that are not trivial to implement at scale. We theorize that increased particle clusters built on spontaneously self-assembled "nucleation sites" expanding with increasing layer count can promote ordered clustering to improve readout quality and sensitivity, having been previously observed in ZnO[51,52],

ZnO/TiO$_2$[53], and CuO nanostructures[54]. Fig. 1a depicts increasingly aligned inter-lattice particle aggregates growing in size and quantity and coinciding with higher layer count. More numerous and larger aligned aggregates will inherently enable more coherent reconstruction of magnetic flux profiles for a dipole emerging at the sample surface compared with randomly aligned NVNPs. This is exemplified by magnetic flux profile of a dipole emerging from the surface of a 100 × 100 µm$^2$ 30 nm thin film of iron oxide measured on 30 layers of NVNPs (Fig. 1b). Fig. 1c provides the corresponding MW spectra of this measured dipole, quantified at 121.5 nT absolute amplitude, and +68.6 nT and -52.9 nT positive and negative amplitudes, respectively (Fig. 1b, rightmost panel), and equivalent to 19.6 MHz difference between spectral dips (Fig. 1c, rightmost panel). We estimate a spectral signal to noise ratio of 0.86 for a randomly aligned NVNP layer (Fig. 1c leftmost panel) and 11.89 for 25% NV centers aligned (Fig. 1c, rightmost panel). These measurements demonstrate detectability of dipoles arising from nano-scale thin films using layered NVNPs and suggest the feasibility of using this architecture for measuring charge movement on similar substrates. To assess this capability, we used an experimental configuration (Fig. 1d) consisting of standard ODMR setup comprising 532 nm laser source applied below the multi-layer NVNP sample with fluorescence collected at the upright position, and a range of microwave energy (2.78 - 2.96 GHz) applied by a printed circuit antenna surrounding the sample. Ensemble particle measurements create characteristic electron spin resonance (ESR) double peak (rightmost panel) from which magnetic field magnitude at a given pixel is calculated with increased fidelity for higher layer count. We turned to leverage this approach for detecting magnetic dipoles of moving charges in microscale conducting elements without requiring specialized fabrication methodologies or materials.

Layering NVNPs at increased layer counts (1, 11, 20 and 30) revealed higher sensitivity and spatial resolution for detecting magnetic dipoles arising from currents applied through a tungsten microwire (~100 µm in diameter) (Fig. 2). Transmitted light images of layered NVNPs show a general upward signal trend corresponding to higher uniformity of the samples over the magnetometer FOV (Fig. 2a, standard deviation was 0.284, 0.262, 0.252 and 0.196 for 1, 11, 20 and 30 layers, respectively). We achieved maximal surface hydrophobicity during manufacturing by applying an isopropanol wash to the NVNP sample, left to dry in ambient temperature between each layer deposition. Similar demonstrations show surface-particle and particle-particle hydrophobic interactions competing with electrostatic attraction to participate in highly uniform nanoparticle self-assembly[45,55]. We next evaluated the ability of this system to detect current-induced magnetic dipoles by applying 33 mA direct current (DC) with either positive or negative polarity (Fig. 2b). We observed a clear dipole reversal during change of current polarity (Fig. 2b, top middle and right panels). Increased layer count coincided with increased spatial resolution and sensitivity (Fig. 2c). For a pixel size of 1 µm$^2$, we found sensitivity (minimum detectable signal) of 89.52 µT, 83.65 µT, 81.51 µT, and 62.64 µT for 1, 11, 20, and 30 layers, respectively. Spatial resolution-dependent sensitivity (Fig. 2c) reveals similar trends, with a slight decrease for 11- and 20-layer samples compared with single layer sample, and a significantly higher performance for 30-layer samples. We find a contrast to RMSE ratio of 4.4856 ± 0.8674, 5.1793 ± 0.8208, 5.4780 ± 0.9084, 6.3654 ± 1.2027 (µ ± σ) for 1, 11, 20, and 30 layers, respectively, supporting this observation (Fig. 2d). Layering shows a plateau in spatial resolution during imaging of current, with a predicted asymptote of 19.096 µm$^2$ (Fig. 2e). This coincides with plateaus observed in irradiated NV bulk crystals with NV layer thicknesses of up to several micrometers.[24–26]

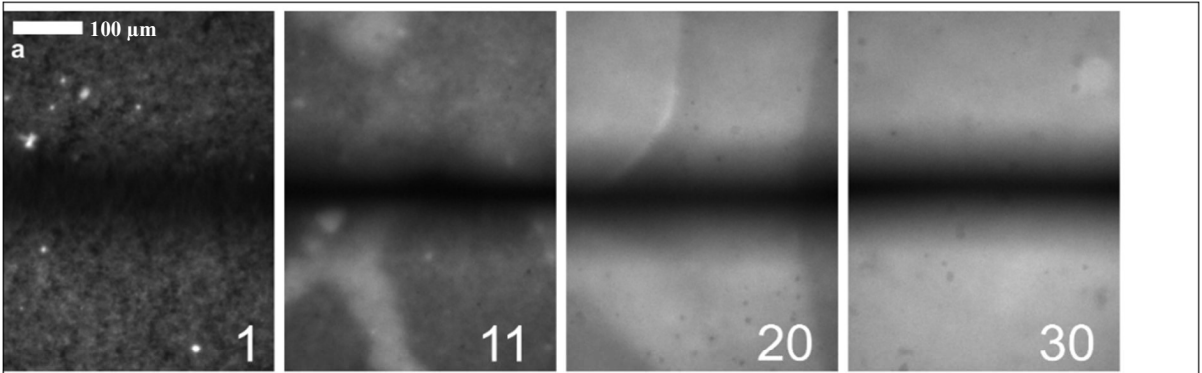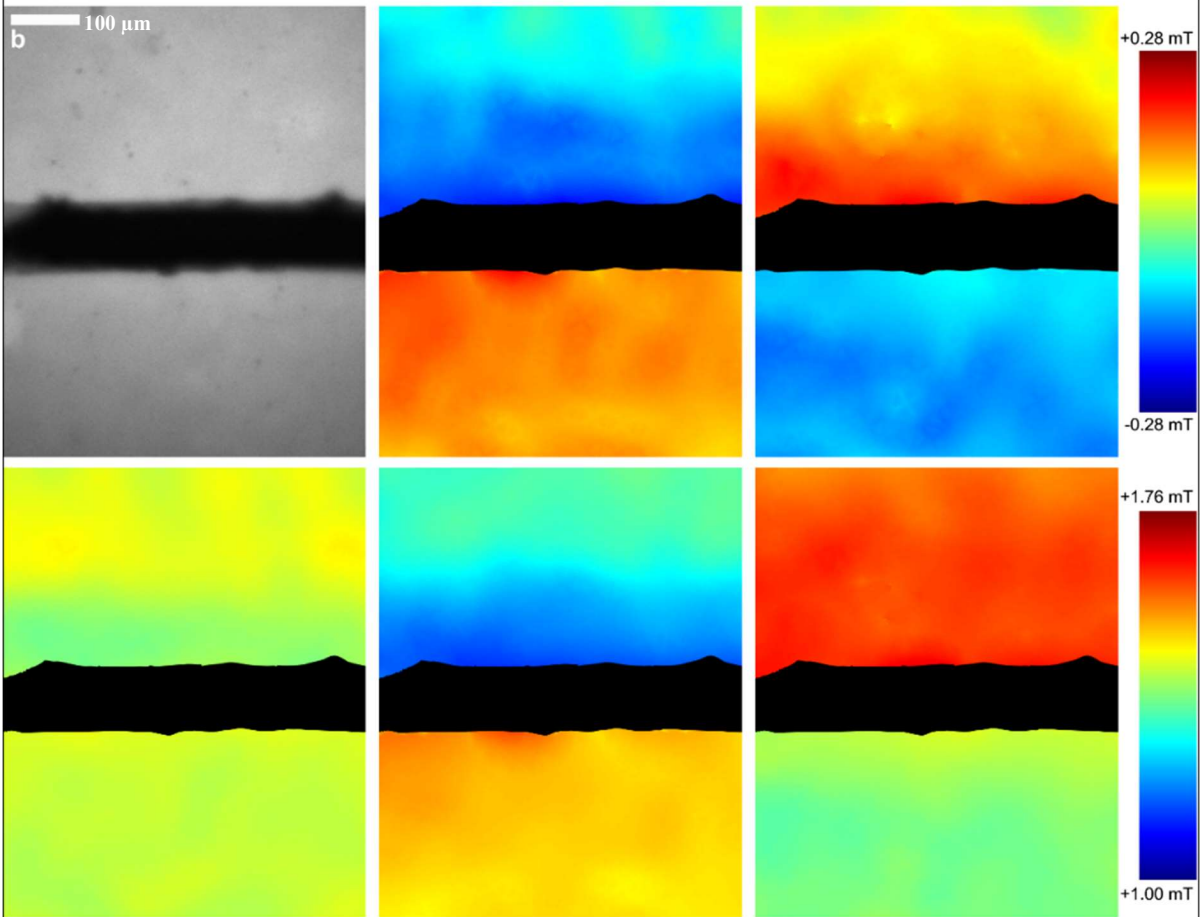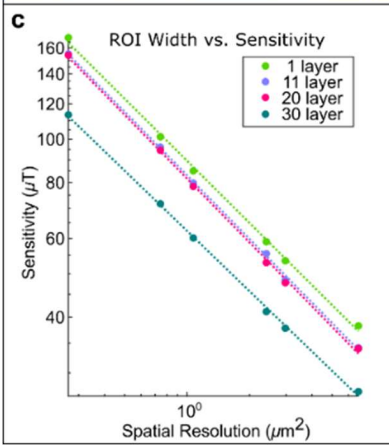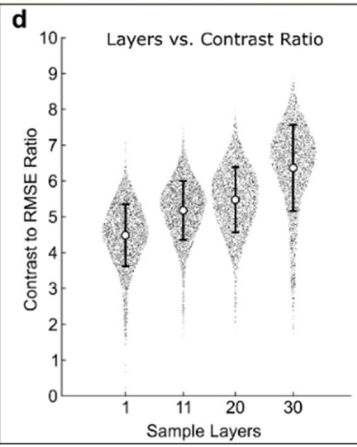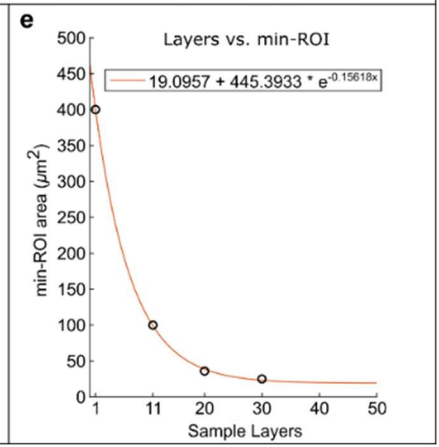

**Fig. 2: Improved sensitivity of NVNP samples with increasing layer strata.** (a) (left to right) 1, 11, 20, and 30 layer samples under 8x magnification. (b) ODMR measurements of magnetic field in response to a DC current of 33 mA through injected through a 100 µm diameter wire. (top left) raw image of wire and 30-layer sample; (bottom left) measurement of magnetic field magnitude with 0 mA current. (bottom middle and bottom right) measurement of field magnitude with 33 mA current applied in opposing polarity. (top middle and top right) subtraction of applied current and zero current measurements to reveal $B_z$-field. (c) Sensitivity, defined as the minimum detectable magnetic field, and its dependence on spatial resolution. Trends show a linear relationship for all layer counts. Performance, defined as the reciprocal slope offset, increases with increased layer count. (d) Ratio of spectral contrast to the RMSE of best-fit curve to data as compared with layer count. (e) Comparison of layer count vs minimum ROI area.

*Magnetic dipole measurements in micropatterned traces*

We turned to testing the ability of layered NVNPs to detect magnetic fields generated by currents in microfabricated devices of various geometries and verified their agreement with finite element predictions (Fig. 3). Two dimensional (2D) ODMR maps were generated during current injection (25-33 mA) in three families of micropatterns: (1) fiducial lines of 64, 32 and 16 µm lateral diameter (Fig. 3a, top left, top middle and top right panel, respectively); (2) single turn inductors with 60, 30 and 15 µm radius (Fig. 3b, top left, top middle and top right panel, respectively); and (3) nine-turn meander inductors with overall length of 150, 120 and 100 µm (Fig. 3c, top left, top middle and top right panel, respectively). For every family, corresponding finite element predictions of magnetic field distribution were performed (Fig. 3a-c, bottom panels corresponding to ODMR data in top panels). Analysis of longitudinal line scans at multiple locations through the devices verify the emergence of magnetic dipole during current injection, with spatial characteristics correlating relatively well with modelling (fig. 3d-f) ), with Pearson product-moment coefficients between ODMR measurements and finite element models given in Table I. We observed a predictable reduction in correlation for larger patterns in all families (for

example, 0.46023 at 50 µm vs 0.88008 at 10 µm for Fiducial Lines, 0.68585 at 200 µm vs 0.94085 at 50 µm for single loop inductors, and 0.18095 at 200 µm vs 0.71371 at 90 µm for meander inductors, leftmost sub-panels in fig. 3d-f, respectively). This could be attributed to larger overall area of Ti/Au traces intruding the collection of NVNP fluorescence within the field of view (FOV). Comparison between the amplitudes of the measured and predicted magnetic dipole for variable geometries (Fig. 3g-i) reveals positive magnetic dipole component (MDC) ranging between +30 µT and +40 µT (predicted: +25 µT to +40 µT) and negative magnetic dipole component ranging between -15 µT and -30 µT (predicted: -30 µT to -35 µT) for fiducial lines (Fig. 3g). Similar quantifications reveal positive MDC of +25 µT to +30 µT (predicted: +20 µT to +30 µT) and negative MDC of -30 µT to -37 µT (predicted: -27 µT to -32 µT) for single turn inductors (Fig. 3h), and +10 µT to +20 µT (predicted: +10 µT to +25 µT) and -15 µT to -25 µT (predicted: -10 µT to -25 µT) for meandering inductor geometries (Fig. 3i). Amplitudes were found to be similar ($p<0.01$) for all small-feature geometries (rightmost panels). This demonstrates the ability of layered NVNPs to detect magnetic field amplitude from micropatterned traces with microscale spatial resolution.

| Device: | Scan 1 (Blue) | Scan 2 (Green) | Scan 3 (Red) |
|---|---|---|---|
| 50um Fiducial | 0.46023 * | 0.56629 * | 0.66336 * |
| 20um Fiducial | 0.95656 ** | 0.94843 ** | 0.90451 ** |
| 10um Fiducial | 0.88008 ** | 0.83348 ** | 0.99412 ** |
| 200um Single Loop | 0.91269 ** | 0.73826 ** | 0.68585 * |
| 100um Single Loop | 0.59843 * | 0.79149 ** | 0.92338 ** |
| 50um Single Loop | 0.91103 ** | 0.92212 ** | 0.94085 ** |
| 300um Inductor | 0.46655 * | 0.42215 * | 0.43593 * |
| 200um Inductor | 0.18095 | 0.24422 | 0.31237 * |
| 90um Inductor | 0.71371 ** | 0.79462 ** | 0.75962 ** |

Table 1: Pearson product-moment coefficients for comparing between ODMR measurements and finite element models

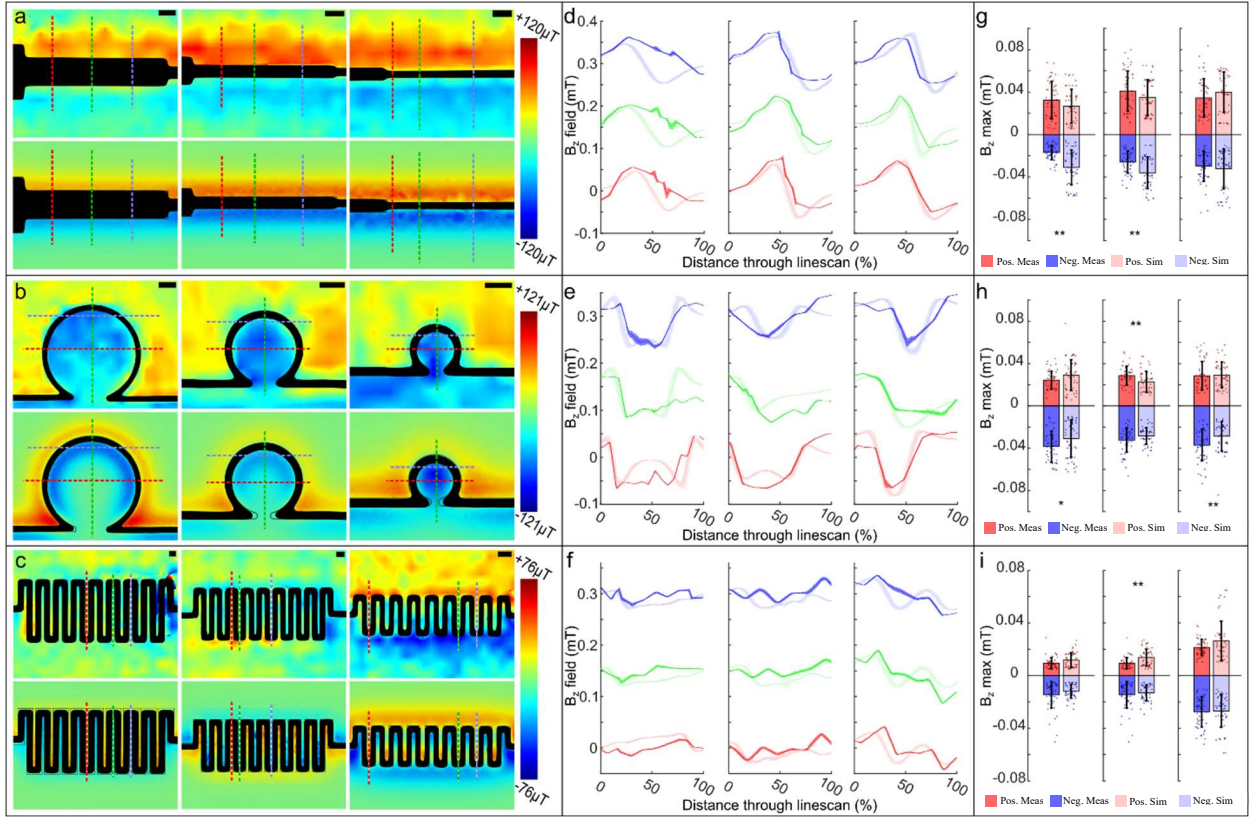

Fig. 3: Direct ODMR measurements and corresponding finite element quantifications of magnetic field amplitude and distribution in variable micro-scale geometries. (a)-(c) $B_z$-field maps of patterns in response to DC current. Top panels show measured $B_z$-field while bottom panels show $B_z$-field predicted by finite element modeling. All scale bars represent 20 μm. (d)-(f) Corresponding plots for $B_z$-field across longitudinal line scans highlighted in panels a-c. Shaded region denotes ± one standard deviation from the mean; dark bars represent ODMR measurements while lighter bars represent simulations. (g)-(i) Comparison of maxima and minima between measured (darker shading) and simulated (lighter shading) positive and magnetic field dipole components. Error bars represent ± one standard deviation from mean (*=p<0.05,**=p<0.01).

*Layering is correlated with inter-cluster particle alignment*

To arrive at a possible mechanistic relationship between magnetic field measurement quality and NVNP layering, we performed analysis of scanning electron microscopy (SEM) images at different layer counts and related them to predicted interparticle alignment that could account for increased performance (Fig. 4). Classifying particles by structural alignment defined as their relative angle of axis revealed an increase in the number of aggregates with preferential directionality of particles as the number of layers increased (Fig. 4a-b). Particles that fell into clusters of similar orientations were identified with same-color overlay (Fig. 4a). To quantify how layering impacted these clusters, mean cluster quality (MCQ) was calculated for each layer count, showing an increase in MCQ with increased layer count (Fig. 4b). Multi-layer samples all showed a marked increase in MCQ compared with a single layer sample, but an insignificant increase between 11 and 20 layer samples, corresponding to trend seen in sensitivity measurements for these same layer counts (Fig. 2c). Based on these aggregation levels, Monte Carlo MW spectra simulations of NVNP samples were generated to determine whether physical preferential orientation of the particles can correspond with the alignment of the NV-centers within the diamond lattice (Fig. 4c). Simulated MW spectral responses were generated with a range of 0% to 100% lattice alignment of particles. Pearson correlation of measured spectral data to these simulations revealed increased correlation to aligned simulations with increased layering, with 11- and 20-layer samples performing similarly (Fig. 4d and Fig. 2c-e). This analysis supports a mechanism whereby increased preferential lattice alignment correlates with increased layer count and could be a contributing factor in the increased sensitivity observed experimentally.

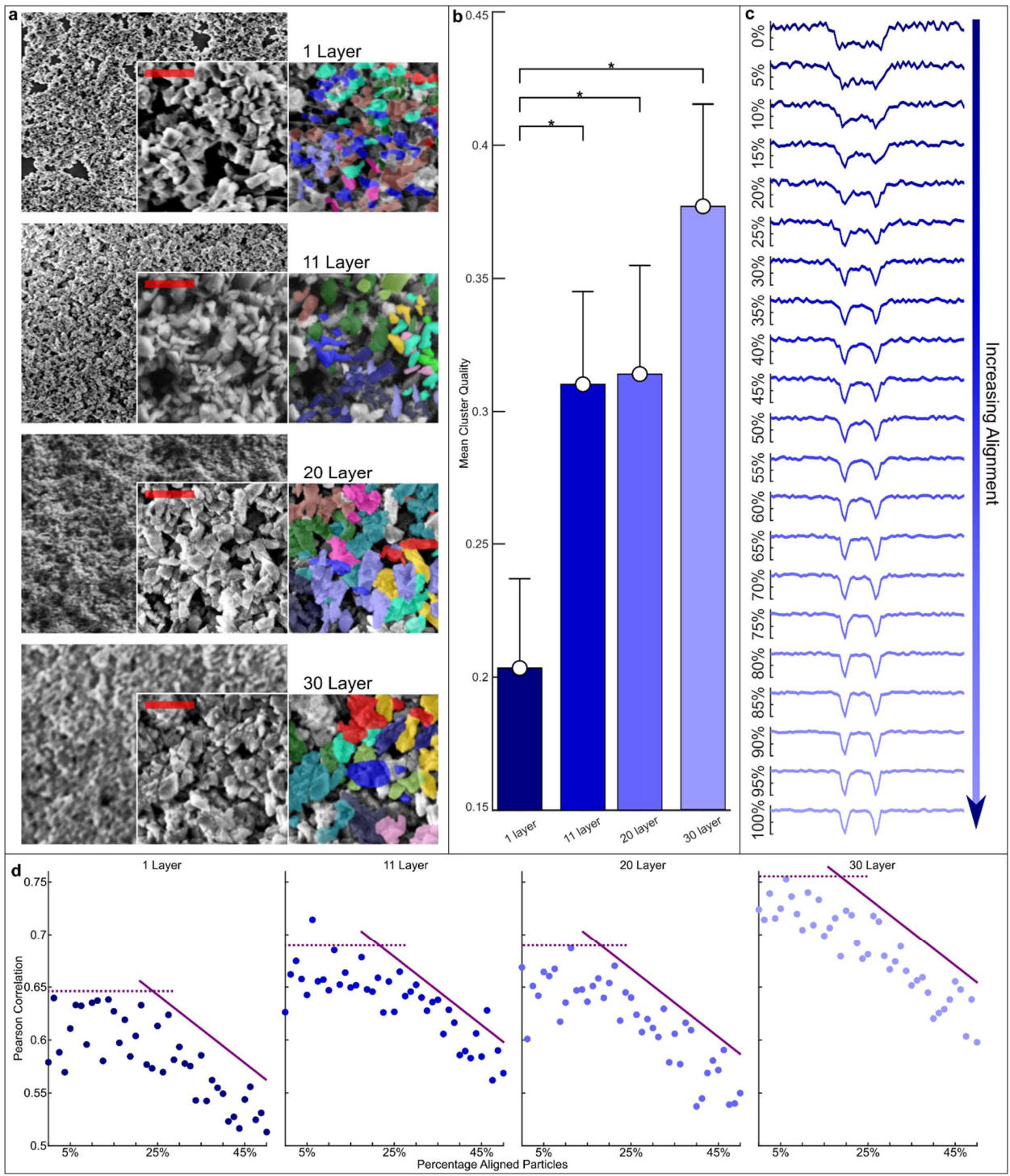

**Fig. 4. Scanning electron microscopy analysis of multilayered NVNPs reveals structural directionality correlating with sensitivity of magnetic field measurements.** (a) Example SEM images of layered particles; Regions of interests (ROIs) with dimensions of 1.5 x 1.5 µm² (insets) were analyzed by segregating single particles and determining their relative angle of axis (right: colored overlay of example ROIs). Scale bar represents 0.5 µm. (b) Number of layers vs. mean cluster quality (MCQ) as a measure of the mean number of same cluster neighbors a given particle exhibited for each cluster. Error bars represent standard deviation. (c) Monte Carlo simulation of spectral response for different layer-counts and for a range of 0% to 100% lattice NVNP alignment. (d) Pearson correlation of measured spectral data to simulation data in panel c for a range of 0 to 50% alignment. Dual estimated asymptotes shown for each condition (dashed and solid purple).

We introduce here a highly accessible micromagnetometry technique that utilizes self-aligned multilayered NVNPs to enable detection of magnetic fields at microscale spatial resolution. We demonstrate the ability of the method to detect fields generated by currents in thin-film micro-patterned electronic elements with implications for the development of new magnetic probes for measuring charge fluctuations *in situ*. A specific and potentially impactful avenue of this technology is using layered NVNPs in biomedical applications for characterizing electromagnetic imaging and recording agents of biophysical events in excitable cells for neuroscience and cardiology[29,56–58]. By avoiding the requirement for fabricating devices directly on diamond or resorting to low-throughput serial scanning of the sample, our method offers an adoptable process for determining device performance and spatial distribution of magnetic field response to inform new and more powerful designs. Other promising efforts demonstrating high throughput controlled deposition of NVNPs use high density bubble printing by direct laser writing of NVNPs that preserve particle spin properties and allow for robust ODMR measurements[59]. Combining this approach and other precision patterning of NVNPs with incremental layering demonstrated here will provide a path to greatly improve applications of nanodiamonds for sensing and catalyze the advent of emerging materials such as magnetoelectric composites for quantum sensing, signal modulation and biological stimulation[60–64].

## Methods

*Micromagnetometry*

A custom-made ODMR magnetometer was built utilizing a flexible PCB MW loop antenna ($d$ = 1 mm, 50 Ω impedance matched) integrated with a 532 nm laser (OBIS 532-80 LS 1264453, Coherent, Santa Clara, CA) and widefield microscope (SM-LUX HL, Leica Biosystems, Wetzlar, Germany). MW signals were generated using an RF signal generator (SG 384, Stanford Research Systems, Sunnyvale, CA) and fed through an RF amplifier (Mini-circuits ZHL-16W-43-S+, Scientific Components Corp., Brooklyn, NY) connected to the antenna. A bias magnetic field was generated using a Halbach array ($k$ = 2) constructed using a neodymium block magnet (BCCC-N52, K&J Magnetics) placed below the layered NVNP sample. Bias field was measured to be 0.9 mT at the sample surface to sufficiently reduce strain and electric field effects and increase magnetic field sensitivity[65,66]. Fluorescence signal changes during current injection were captured using an upright microscope (SM-LUX HL, Leica Biosystems, Wetzlar, Germany) mounted with a CMOS camera (CS165MU1, Thorlabs, Inc., Newton, NJ) operating at 12 frames/sec and a resolution of 720 × 540 pixels with a corresponding region of interest (ROI) size of 527 × 395 µm. A total of 181 frames surrounding resonance at ~2.87 GHz were acquired while sweeping between 2.78 and 2.96 GHz at 1 MHz intervals for a total of 181 data points per pixel and an acquisition time of 15 min. The image capture and delivery of microwaves and lasers were directly controlled through a MATLAB (MathWorks, Inc. Natick, MA, USA) interface and in-house MATLAB routines.

*Preparation of layered NVNP samples*

To prepare layered NVNP samples, we delineated a 5x5 mm² region at the center of micro-glass wafers (thickness 160-190 µm) and applied the region with Isopropyl alcohol (IPA) that was then left to evaporate to form a hydrophobic region. This step was repeated three times. We then applied 2 µL of 1 mg/mL fluorescent nano-diamonds (Cat #: 798088, Sigma-Aldrich, St Louis, MO) onto the center of the hydrophobic region and allowed the solution to dry in room temperature (RT) for 2 hours to allow NVNPs to settle. These steps were repeated to achieve the desired number of layers. For samples used in measuring field distributions at different microwire arrangements, we used 100 µm diameter tungsten wire affixed to a 2" × 3" glass slide that was faced down onto the layered NVNP sample with wired centered within the microscope FOV. For all samples, NVNPs were measured at average size of 130 nm using dynamic light scattering and profilometry was carried out on layered samples (DEP KLA-Tencor P7 Profilometer (KLA Corporation, Milpitas CA).

*Device microfabrication*

Standard photolithography was used to fabricate three families of microdevices. The first contained a series of fiducial wires with stepwise decreases in trace width intended to isolate the effect of geometry width on field detectability. The second was a series of single turn coils with varying diameters to elucidate magnetic field distribution around circular loops. The third was a series of meandering inductors with differing lengths and trace widths. These served as a method for evaluating the effect of tight turn-to-distance ratios on magnetic field spread. Glass substrates were spin-coated with S183 photoresist (30 s, 3000 RPM, 1.3 µm thickness) baked at 110 °C for 1 min, followed by soft contact lithography (Karl Suss MA6, 15 s exposure time, 10 mW/cm² broadband mercury lamp) to pattern traces. Samples were developed in MF-321 (Kayaku Advanced Materials, Inc., Westborough MA) for 60 sec, washed with deionized (DI) water and dried with $N_2$. A Ti/Au bilayer (10/100 nm, respectively) was evaporated using e-beam deposition and samples were then rinsed with IPA and DI water and dried with $N_2$ and lifted off in an ultrasonic bath at a medium-high vibration rate with acetone. This was followed by a 10 sec oxygen plasma treatment of the descum surface (YES R3 Plasma Asher, 250 W, 80 SCCM O2),

*Finite Element Simulations*

Magnetic field profiles were simulated using COMSOL Multiphysics 5.5 (COMSOL Inc., Stockholm, Sweden). Computer aided designs representing respective microfabricated geometries were imported into COMSOL and the Electric Currents (*ec*) and Magnetic Field (*mf*) interfaces of the AC/DC Module. Input current was applied between two ports at opposite edges of the simulation arena and resulting magnetic field profiles were quantified at planes 50 μm above the device surface as the *z*-component of magnetic flux density (B-field) reflecting ODMR readouts. Field maps obtained were exported to MATLAB (MathWorks, Inc., Natick, MA) for further processing and analysis. Maximum and minimum magnetic field from each empirical dataset was compared to its respective simulation dataset.

*Scanning Electron Microscopy Analysis*

Scanning electron microscopy (SEM) analysis was performed on NVNP samples prepared on $Si/SiO_2$ wafers using the same methods for ODMR measurements. Sets of images were taken of 1, 11, 20 and 30 layers and imported to the open source image processing software Fiji[67] for analysis. Each particle was outlined, and a set of random ROIs was generated for each layer count with the orientation (directionality) of each particle determined by the angle of its long axis. Clusters were then defined by assigning a particle to a cluster if and only if its angle was within tolerance level defined as the mean orientation of the cluster and less than 2.3 μm from another particle that was already part of that cluster. Cluster quality (CQ) was calculated to determine the closeness of particles within a cluster with the mean number of same cluster neighbors, with mean cluster quality – MCQ defined as:

$$MCQ = \mu_{unique}\left(\bigcup_0^n d_{particle,n} < d_{particle,max}\right) \quad (1)$$

Where $d$ is the interparticle distance and $\mu$ is a binary coefficient representing a unique particle pair. A simulation of optical response from NV centers was performed using in-house MATLAB routines to further study observations within SEM analysis. Particles were assumed to have an equal proportion of NVs in all four crystallographic orientations and average size measured at 130 nm using dynamic light scattering. Magnetic field was defined, and a corresponding reference NV axis vector system was created within MATLAB to represent a tetrahedral geometry with a user-defined number of orientations within the field of view. Random orientations for NV geometries were generated by utilizing a three-dimensional rotation matrix. For each orientation, the yaw, pitch, and roll for the rotation matrix was randomly drawn from a uniform distribution with support (0,1). Random trials were reproducible by setting MATLABs random number generator seed to 1. Each reference NV vector was then multiplied by the rotation matrix to create a new random NV orientation. The component of the arbitrary magnetic field was determined along each NV axis for a given orientation, and the fluorescence response was calculated. Individual responses were then summed to calculate the overall simulated response from multiple particles. To evaluate the effect of particle lattice alignment, an alignment factor was defined as the fraction of particle lattices aligned with one another and tested for 0-100% alignment in steps of 5%. Measured spectral data was correlated with simulation data for ranges of 0-50% in steps of 5% (Pearson correlation).

## Acknowledgements

This work was supported by the National Institute of Neurological Disorders and Stroke and the Office of the Director's Common Fund at the National Institutes of Health (Grant DP2NS122605 to AH) and the National Institute of Biomedical Imaging and Bioengineering (Grant K01EB027184 to AH). This material is also based on research supported by the US Office of Naval Research under award numbers N00014-23–1-2006 and N00014-22–1-2371 to A.H. through Dr. Timothy Bentley and the Wisconsin Alumni Research Foundation (WARF). The authors gratefully acknowledge the use of facilities and instrumentation supported by NSF through the University of Wisconsin Materials Research Science and Engineering Center (DMR-1720415). We thank Dr. Shlomo Kolkowitz for advice on ODMR and for supplying wire antenna.